\documentclass[conference]{IEEEtran}
\IEEEoverridecommandlockouts
\usepackage{cite}
\usepackage{amsmath,amssymb,amsfonts}
\usepackage{graphicx}
\usepackage{textcomp}
\usepackage{xcolor}
\usepackage{siunitx}
\DeclareSIUnit \voltampere { VA } 
\DeclareSIUnit \var { var } 
\graphicspath{{./Images/}}

\usepackage[linesnumbered,ruled]{algorithm2e}
\begin{document}
%
\title{On Feasibility and Flexibility Operating Regions of Virtual Power Plants and TSO/DSO Interfaces}

\author{
\IEEEauthorblockN{Shariq Riaz and Pierluigi Mancarella}
\IEEEauthorblockA{Department of Electrical and Electronic Engineering, The University of Melbourne,
Australia\\
\{shariq.riaz,pierluigi.mancarella\}@unimelb.edu.au 
}
\thanks{The authors gratefully acknowledge the partial support for this research received from the Victorian Government’s \emph{veski} programme.}
}


\maketitle

\begin{abstract}
Distributed energy resources are an ideal candidate for the provision of additional flexibility required by the power system to support the increasing penetration of renewable energy sources. The integrating large number of resources in the existing market structure, particularly in the light of providing flexibility services, is envisioned through the concept of virtual power plant (VPP). To this end, it is crucial to establish a clear methodology for VPP flexibility modelling. In this context, this paper first puts forward the need to clarify the difference between \emph{feasibility} and \emph{flexibility} potential of a VPP, and then propose a methodology for the evaluation of relevant operating regions. Similar concepts can also be used to modelling TSO/DSO interface operation. Several case studies are designed to reflect the distinct information conveyed by feasibility and flexibility operating regions in the presence of “slow” and “fast” responding resources for a VPP partaking in the provision of energy and grid support services. The results also highlight the impact of flexible load and importantly network topology on the VPP feasibility (FOR) and flexibility (FXOR) operating regions.

\end{abstract}



\section{Introduction}
In the pursuit of energy system decarbonisation, replacement of conventional thermal power plants with variable and partly unpredictable renewable energy resources (RES) has created new challenges in power systems, primarily due to increased flexibility requirements and simultaneous reduction of available operational flexibility mainly provided by conventional power plants. On the other hand, increasing electricity prices and decreasing technology cost have boosted investments in distributed household photovoltaic (PV) and batteries systems, coupled with the development of smart grid technologies that allow better control over distributed energy resources (DER). This creates new opportunities for flexibility provision from DER, particularly for the purpose of frequency control ancillary services (FCAS) that are essential for stability and security in low-inertia power systems~\cite{Gonzalez2018}.

Currently, aggregated flexibility from DER and various demand-side resources that can provide Demand Response (DR) is still vastly an untapped resource, restrained by the traditional power system management approach of treating distribution networks as passive entities~\cite{Pudjianto2007,Capitanescu2018,Gonzalez2018,Papavasiliou2018}. Nonetheless, significant benefits can be derived through proper management and control of DER as they have the potential to assume a central role in the future grid stability and security~\cite{Riaz2017c}. Recently, in the Open Energy Network (OEN) consultation paper~\cite{AEMO18}, Energy Network Australia and the Australian Electricity Market Operator expressed the need and urgency to integrate DER and tap their flexibility potential in order to reduce electricity costs. The OEN paper also explores the impact from collective orchestration of DER to deliver a significant amount of operational flexibility and potential to influence grid management strategies.

However, the key challenge lies in finding ways to facilitate effective integration of a vast number of devices in the existing market structures. Central to the aggregation and coordinated control of DER is the concept of virtual power plant (VPP)~\cite{Pudjianto2007}, which seeks to address the challenge by clustering a large number of devices, backed by appropriate control policies, to efficiently deliver some of the key grid support services in an economically viable manner. A VPP takes into account operational constraints of DER coupled with network restrictions and offers the aggregated capacity in various markets for the provision of energy and grid support services, thus boosting business case of grid participation of DER~\cite{Pudjianto2007}. In line with VPP flexibility modelling, recently there is rising interest in the ideas of flexibility estimation at transmission system operator (TSO) and distribution system operator (DSO) interface boundary of an active distribution network~\cite{Pudjianto2007, Heleno2015,Silva2018, Gonzalez2018, Saint-Pierre2016, Mancarella2013J, Contreras2018,Capitanescu2018}\footnote{Also to note is various work that deals with different integration approach for VPP into various markets, tackling uncertainty from DER, incorporating multi-energy storage systems, and proposing real-time control, uncertainty management, active monitoring and risk minimization algorithms~\cite{Kardakos2016,Mashhour2011a,Mnatsakanyan2015,Giuntoli2013,Wang2016,Dabbagh2016,Ruiz2009}.}. 

The concept and high-level structure of a VPP were comprehensively discussed~\cite{Pudjianto2007} in the FENIX project, classifying VPPs into commercial and technical roles and establishes simple algorithms for estimating their static characteristics. A general framework for quantifying and techniques for visualizing the operational flexibility from generic resources is described in~\cite{Ulbig2015}: this work deploys Minkowski summation technique that yields a good estimation of the aggregated flexibility in the absence of network constraints. The relationship between power consumption and voltage is exploited in~\cite{Capitanescu2018} through the control of on-load tap changer at the interface of DSO and TSO. In~\cite{Heleno2015,Silva2018} cost maps are proposed as an instrument to provide flexibility information to TSO. Time dependency and the uncertainty of feasible operating space are detailed in~\cite{Gonzalez2018,Saint-Pierre2016}, while techno-economic modelling of flexibility from multi-energy vectors is introduced in~\cite{Mancarella2013J} and a linearisation approach to reduce the computation time of VPP non-linear frameworks is elaborated in~\cite{Contreras2018}.

\begin{figure}[!t]
	\centering
	\includegraphics[width=85mm] {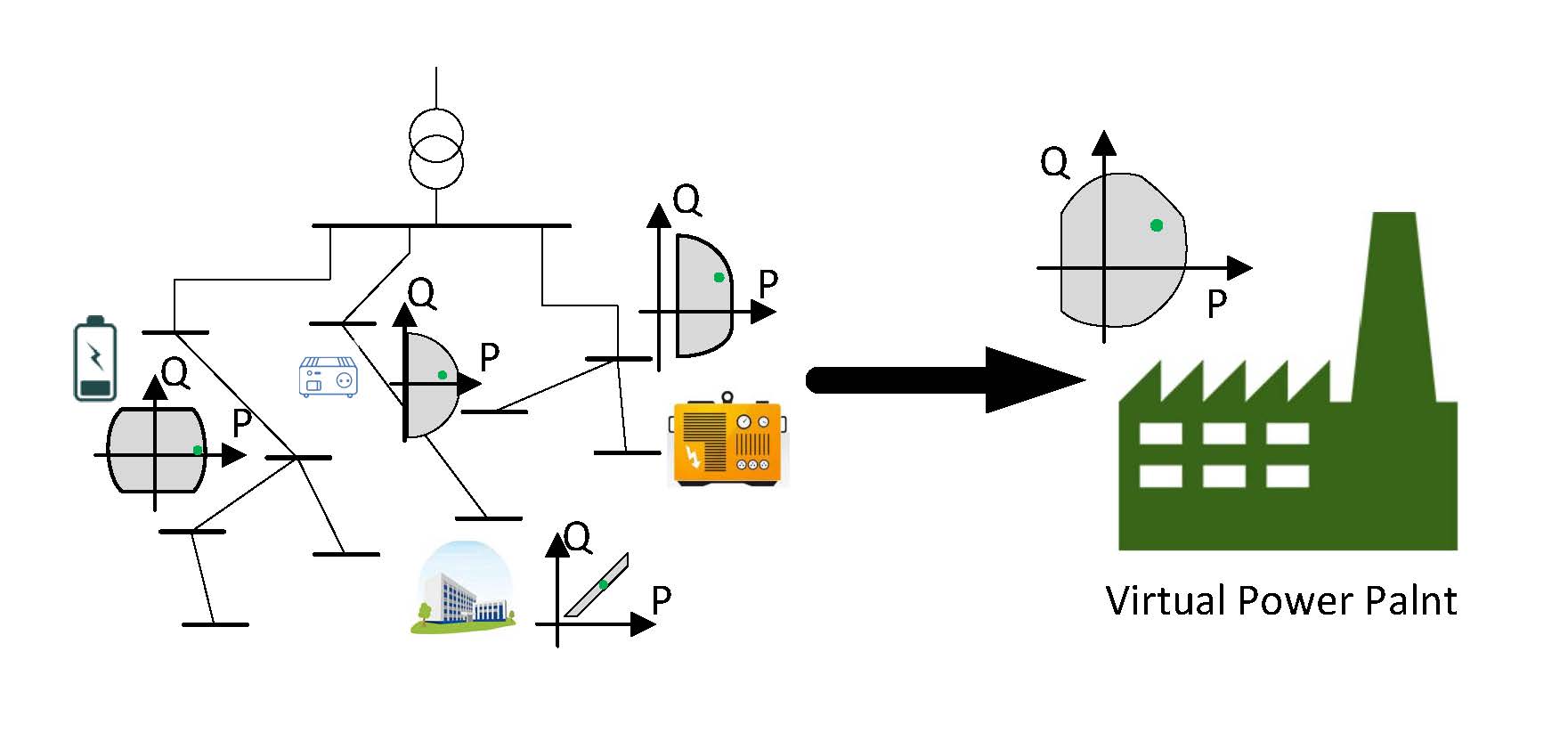}
	\vspace{-.5em}
	\caption{Conceptual visualisation of a virtual power plant’s \emph{feasibility} operating region.}
	\vspace{-1em}
	\label{fig:VPP}
\end{figure}

Almost all of these works are primarily focused on the estimation of operational feasibility and uses the concept of “flexibility” interchangeably with “feasibility”. However, this is only valid if all of the considered resources have a sub-second response time, for example, converter base systems and some demand response (DR) technologies. In contrast, in the context of the provision of fast and slow responding resources and relevant market services, “feasibility” is not sufficient to model the actual “flexibility” characteristics of a VPP. Systematic evaluation of the temporal ability to move between two set-points (“flexibility”) that are located in a feasible operating region (“feasibility”) is therefore needed to assess the VPP potential to optimise its participation in energy and grid service provision. The conceptual visualisation of the “feasibility operating region (FOR)” of a VPP corresponding to a given operation set-point is represented in Fig.~\ref{fig:VPP},
while “flexibility operating regions (FXOR)” would be a subset, as elaborated on below.

In light of the above, as various works treat flexibility synonymously to feasibility, there is a need for precise definitions and identifying the distinct role of flexibility apart from feasibility. Thus, the key objectives and contributions of this paper are as follows:
\begin{itemize}
	\item Acknowledge the distinct information provided by “feasibility” and “flexibility” in the context of a VPP;
	\item Methodology for the assessment of VPP’s flexibility potential, as a subset of its feasible operating set;
	\item Identify the role of the information provided by feasibility and flexibility for a VPP providing different grid services.
\end{itemize} 

The proposed methodology is demonstrated on the IEEE 33 bus network. The case studies are designed to bring out the usefulness of clearly distinguishing feasibility and flexibility, with application in the Australian context. Furthermore, case studies also explore the impacts of flexible loads and network topology on the FOR and FXOR of a VPP.

The remainder of the paper is structured as follows: Section~\ref{Meth} discusses the state of art on the topic under analysis and outlines the methodology adopted in this work. Section~\ref{SS} introduces the test system, cases and assumptions used in the simulations, while the results from the case studies are then discussed in Section~\ref{RaD}. Section~\ref{Cnl} concludes the paper.  
 
\section{Methodology} \label{Meth}
This Section provides key definitions, discusses the state of the art, and outlines the methodology adopted in this paper.

\subsection{Key Definitions}
As acknowledged by~\cite{Gonzalez2018}, there is a need for developing a consistent terminology. Some definitions in the context of presented work are therefore described as follows:
\begin{enumerate}
	\item \textbf{\textit{Dispatch Power ($S^{\lambda}$)}:} is defined as the complex power (pair of active and reactive powers) exchange between VPP and the grid, resulting from the VPP participation in the energy market alone.
	\item \textbf{\textit{Feasible operating Region (FOR)}:} represents the set of all feasible dispatch power points of a VPP, mathematically represented as $\cup S^{\lambda}$. In general, this is time-varying and depends on the current dispatch point too.
	\item \textbf{\textit{Ancillary Power  ($\Delta S^\text{a}$)}:} is the complex power deviation of a VPP from a given dispatch power point to participate in grid support services.	
	\item \textit{\textbf{Flexibility}:} is the measure of VPP’s responsiveness to deviated from its dispatch power point; in other words, it represents how fast a combination of resources within the VPP can be deployed to provide grid support services. 
	\item \textit{\textbf{5)	Flexibility Operating Region (FXOR)}:} ): is a quantitative representation of the VPP flexibility; it consists on the set of achievable \textcolor{black}{ancillary power} points in a given amount of time ($\tau$) and mathematically can be represented as $S^{\lambda} + \cup \Delta S^\text{a}(t)|_{t\leq\tau}$.
\end{enumerate}

\subsection{State of the art} \label{SoA}
There are two fundamental methodologies described in the literature to estimate the feasibility potential of a VPP, namely, i) Monte Carlo estimation~\cite{Heleno2015,Gonzalez2018} and ii) optimization approach~\cite{Ulbig2015,Contreras2018,Silva2018,Capitanescu2018}. Both methodological approaches have the potential to be extended in the context of estimating flexibility potential.

In Monte Carlo based simulations, first, a large number of power flow instances are executed for randomly generated operating points; then, non-convergent points are discarded, and the remaining points provides the feasibility region of a VPP. While this method is simple and can generally be applied to various network topologies, it is inherently limited by the fact that a large number of operating points are required to get an accurate estimate.

In contrast to the Monte Carlo approach, the optimization approach directly aims to find the boundary of the feasibility region by considering network and DER operation constraints. In this method, first, four optimization problems are setup aiming to find the minimum and maximum active and reactive power requirement of the VPP~\cite{Contreras2018,Silva2018,Capitanescu2018}. This provides with four extreme points of the PQ capability curve of a VPP. Then, the boundary is further refined by solving the optimization problem for intermediate points. This approach is more suitable for small to medium scale networks; however, the size and computation complexity grows exponentially with network size and number of DER. Also, due to network constraints, the optimization problem is non-linear in nature, though promising research is ongoing to propose linearisation of network and DER constraints which might be suitable to reduce the computational complexity of this method (e.g.,~\cite{Contreras2018}). 

Since the main aim of this paper is to distinguish between the feasibility and flexibility potential of a VPP, a Monte Carlo approach, which is simple and straightforward, is deployed for both feasibility and flexibility estimations.
 
\subsection{Feasibility Estimation}
The procedure for assessment of FOR is summarised in Algorithm~\ref{Algo:FOR}, derived from the method provided in~\cite{Gonzalez2018,Heleno2015}.
\begin{algorithm}[t] \label{Algo:FOR}
	\SetKwInOut{Input}{Input}	
	\Input{Network constraint, loads and DER operational constraints}
	Randomly generate correlated DER operating points   $n\in \mathcal{N}$, within each DER operational constraints\;
	\For{$n\gets1,\left| \mathcal{N}\right| $}{
		Solve power flow\;
		\eIf{Network constraint are violated}
		{
			Discard $n$\;
		}
		{
			Add $n$ to $\mathcal{O}$\;
			Add corresponding VPP dispatched point to FOR\;
		}
		
		}
	\SetKwInOut{Output}{Output}
	\Output{FOR and $\mathcal{O}$}
	\caption{Algorithm for determining FOR}
\end{algorithm}
 The inputs to the algorithm are network constraints, load requirement of each bus and operational limitations of participating resources in the VPP arrangement. The algorithm starts by generating a large number of vectors ($n_{1\times 2R}$) consisting of operational points for the resources:
 \begin{equation*}
 n=\{P_1,P_2,\dots,P_R,Q_1,Q_2,\dots,Q_R\},
 \end{equation*}
where, R is the total number of resources in the VPP. Then, an AC power flow is solved for each combination of operating points $n$ and the solution is checked against network constraints, that is bus voltage and component thermal limits. The points satisfying network limits are then included in a set of feasible operating points $\mathcal{O}$ and the resulting dispatched point of VPP $S^{\lambda}$ is added to the FOR, otherwise, the point is discarded. In the end, the algorithm returns the FOR along with the associated set of DER operating points. The FOR returned at the end of Algorithm~\ref{Algo:FOR} is basically the aggregated PQ capability curve of a VPP and contains sufficient information the VPP to participate in the energy market.

\subsection{Flexibility Estimation}
Flexibility reflects the time required by a VPP to deviate from a particular dispatched point to another point within the FOR. It depends upon the activation time, ramp rates and status of the participating DER. The process to identify the achievable operating points from the dispatch point in a given amount of time is outlined in Algorithm~\ref{Algo:FLX}.
\begin{algorithm}[t] \label{Algo:FLX}
	\SetKwInOut{Input}{Input}	
	\Input{Outputs of Algorithm~\ref{Algo:FOR}, VPP market dispatched point ($S^{\lambda}$), resources activation time, ramp rates and target time intervals ($\tau$)}
	\For{$n\gets1,\left| \mathcal{O}\right| $}{
		Calculate time ($t$) required by resources to change their output to $n$ from set-points corresponding to market dispatch\;
		\If{$t\leq\tau$}
		{
			Add corresponding deviated dispatched point to flexibility curve $\mathcal{X}_\tau$\;
		}		
	}
	\SetKwInOut{Output}{Output}
	\Output{$\mathcal{X}_\tau$}	
	\caption{Algorithm for determining the FXOR}
\end{algorithm}
The activation time, ramp rates, status and operating points of resources corresponding to FOR, dispatch point of the VPP, and target time intervals ($\tau$) serve as the inputs for the algorithm. The algorithms calculate the time required to change the output of VPP from the current dispatch point to all other points within the FOR. Feasible points that can be achieved within the specified time interval are included in the flexibility curve ($\mathcal{X}_\tau$) that bounds the FXOR, which is returned by the algorithm. The FXOR is a reflection of VPP’s ability to participate in FCAS markets and providing reactive power services after being dispatched at a particular dispatch point in the energy market.

\section{Simulation Setup} \label{SS}
Description of the test system, study cases and assumptions are presented in this section.
\subsection{Test System} \label{TS}
\begin{figure}[!t]
	\centering
	\includegraphics[width=50mm] {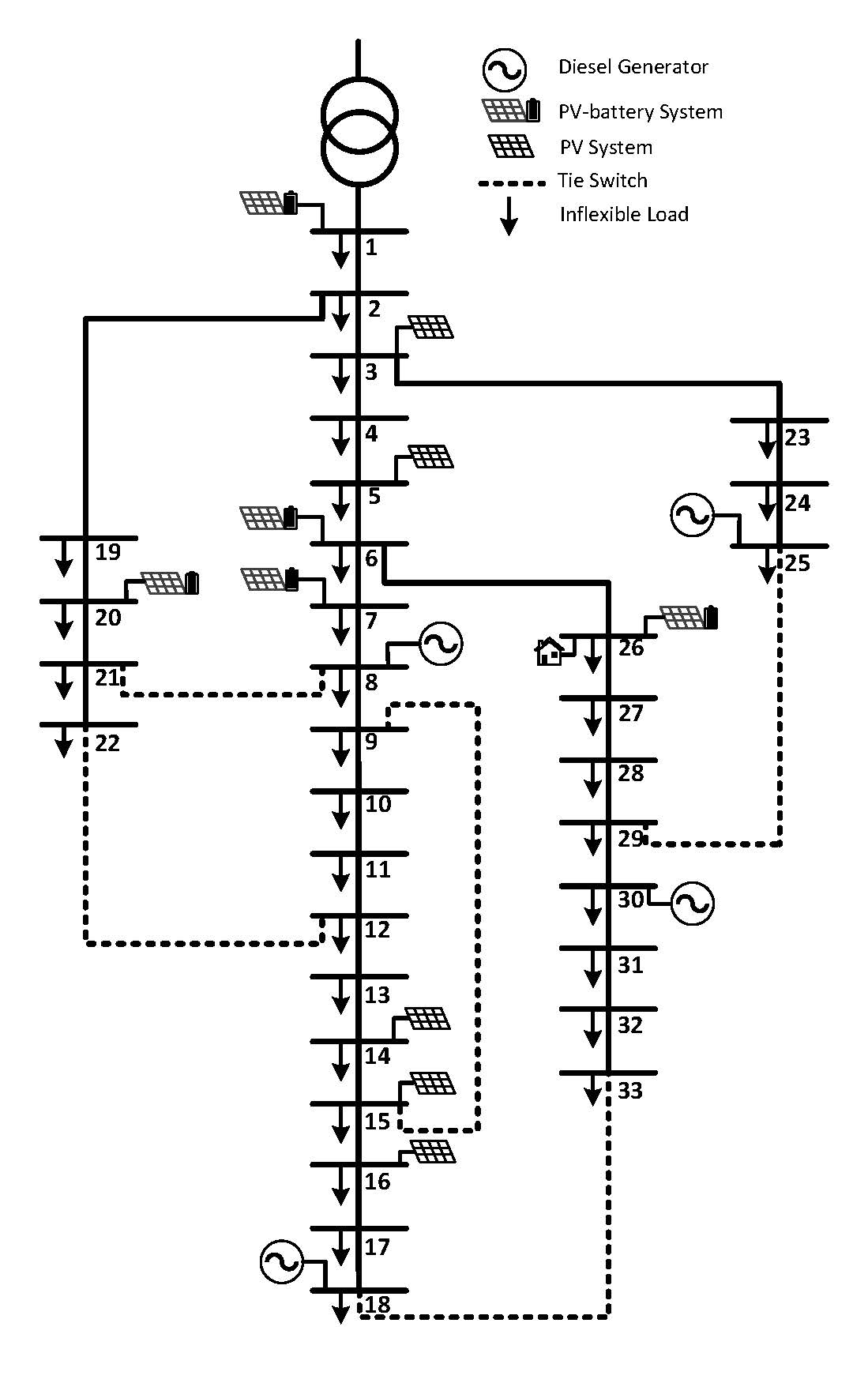}
	\caption{Modified IEEE 33 bus network.}
	\label{fig:Network}
\end{figure}

\begin{figure}[!t]
	\centering
	\includegraphics[width=85mm] {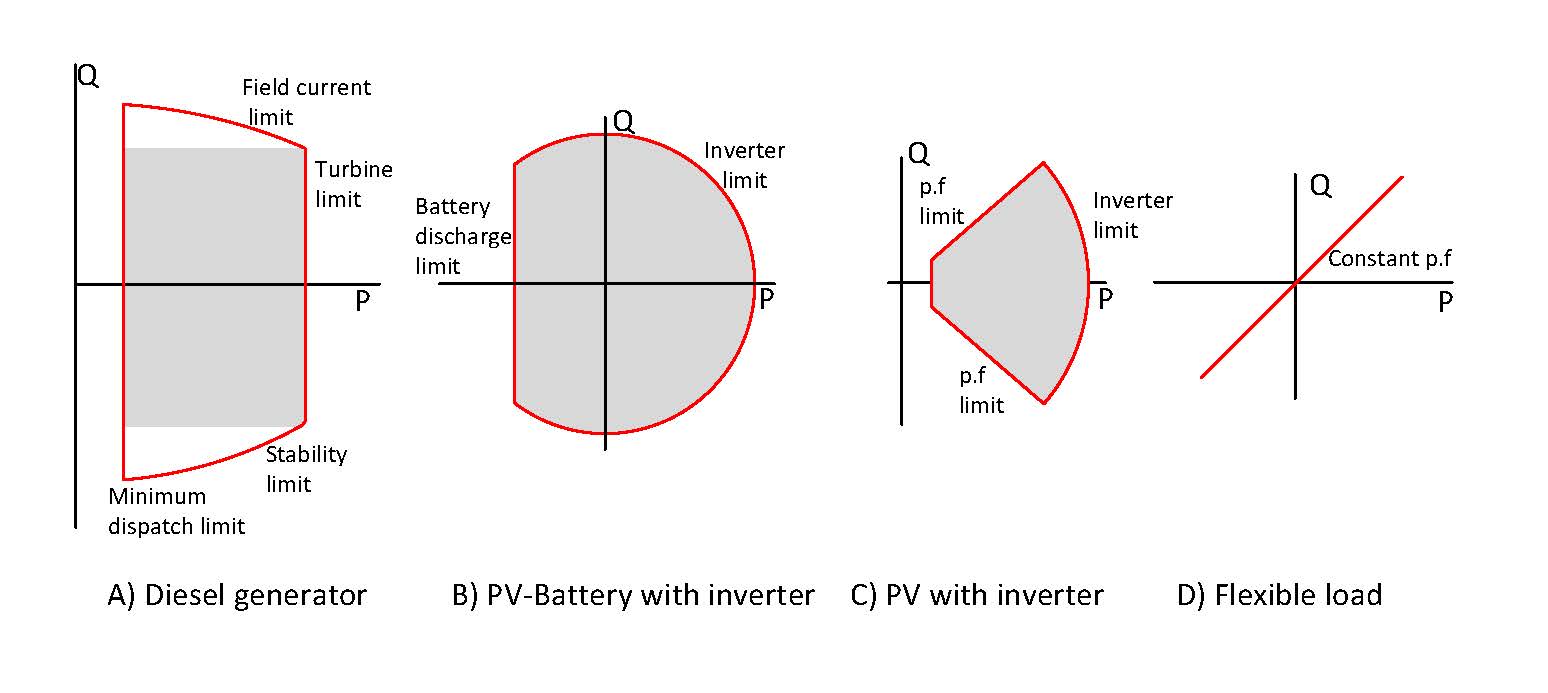}
	\caption{Actual (red) and assumed (grey) PQ capabilities of resources considered in this study.}
	\vspace{-2em}
	\label{fig:RscPQ}
\end{figure}

The presented studies use a modified IEEE 33 bus system as a test bench to distinguish between feasibility and flexibility, as shown in Fig.~\ref{fig:Network}.
Network characteristics such as line impedances and load requirements are taken from the original model first presented in~\cite{Baran1989}. The network is then populated with four diesel generators, five buses with aggregated rooftop-PV systems ranging from \SI{28}{\kilo\watt} to \SI{38}{\kilo\watt} summing to a total installed capacity of \SI{162}{\kilo\watt}; and five buses with household PV-battery systems with the average capacity of \SI{38}{\kilo\watt} of PV backed by \SI{20}{\kilo\watt\hour} of storage. Therefore, the total installed capacity of diesel generators, PVs and batteries in the system is \SI{1}{\mega\watt}, \SI{352}{\kilo\watt} and \SI{100}{\kilo\watt}, respectively. The net active and reactive power demand of the system is \SI{3.71}{\mega\watt} and \SI{1.76}{\mega\var}, respectively. Specification of installed diesel generators are given in Table~\ref{Tbl:Rcs},
\begin{table} [!t]
	\renewcommand{\arraystretch}{1.0}
	\centering	
	\caption{Diesel generator's specifications.}
	\label{Tbl:Rcs}
	\begin{tabular}{|c|c|c|c|c|c|c|}
		\hline Bus & \multicolumn{4}{c|}{Power limits} & \multicolumn{2}{c|}{Time}  \\ \cline{2-7}  
		& ${P}_{min}$ & ${P}_{max}$ & ${Q}_{min}$ & ${Q}_{max}$ & Activation & Ramp \\ 
		& (\SI{}{\kilo\watt}) & (\SI{}{\kilo\watt}) &(\SI{}{\kilo\var}) &(\SI{}{\kilo\var}) &(\SI{}{\sec})  & (\SI{}{\sec}) \\ \hline
		8,30 & 15 & 100 & -40 & 60 &15 & 20 \\ \hline
		18 & 150 & 500 & -200 & 300 &40 & 60 \\ \hline
		25 & 60 & 300 & -120 & 180 &25 & 45 \\ \hline
	\end{tabular}
\end{table}
 where as the models of deployed resources are shown in Fig.~\ref{fig:RscPQ}.
PV, battery and flexible loads are subjected to activation and ramp times between \SI{0.1} and \SI{0.3} seconds, so these resources possess the capability to move from minimum to maximum values of power within a second and \emph{vice versa}. Furthermore, PV inverters are subjected to a power factor limit of \SI{0.9} and the minimum output power is set to \SI{10}{\percent} of the inverter rated value. The FOR and FXOR are calculated for one instant of time and do not consider the inter-temporal constraints. Moreover, it is assumed that the VPP is equipped with proper communication, control and incentive mechanism and is authorised to operate resources within their PQ capabilities while acknowledging their operational requirements.
\subsection{Test Cases}
Three different study cases are designed to discriminate between feasibility and flexibility potential of a VPP, along with the impact of flexible demand and network restrictions. More specifically, \textbf{Case I} first deploys test system explained in Section~\ref{TS} and highlights the usefulness of flexibility in the context of a VPP partaking in grid support services. Second, \textbf{Case II} considers \SI{5}{\percent} of the load at each bus to be flexible, in order to study its impact on the VPP's flexibility. Finally, in \textbf{Case III} tie switches connecting bus 8 to 21 and 12 to 22 are open to study the impact of network reconfiguration.  
 
\section{Results and Discussion} \label{RaD}
The results of the feasibility and flexibility potential of DER aggregation in the proposed study cases are presented as follows.
\subsection{Feasibility}
\begin{figure}[!t]
	\centering
	\includegraphics[width=80mm] {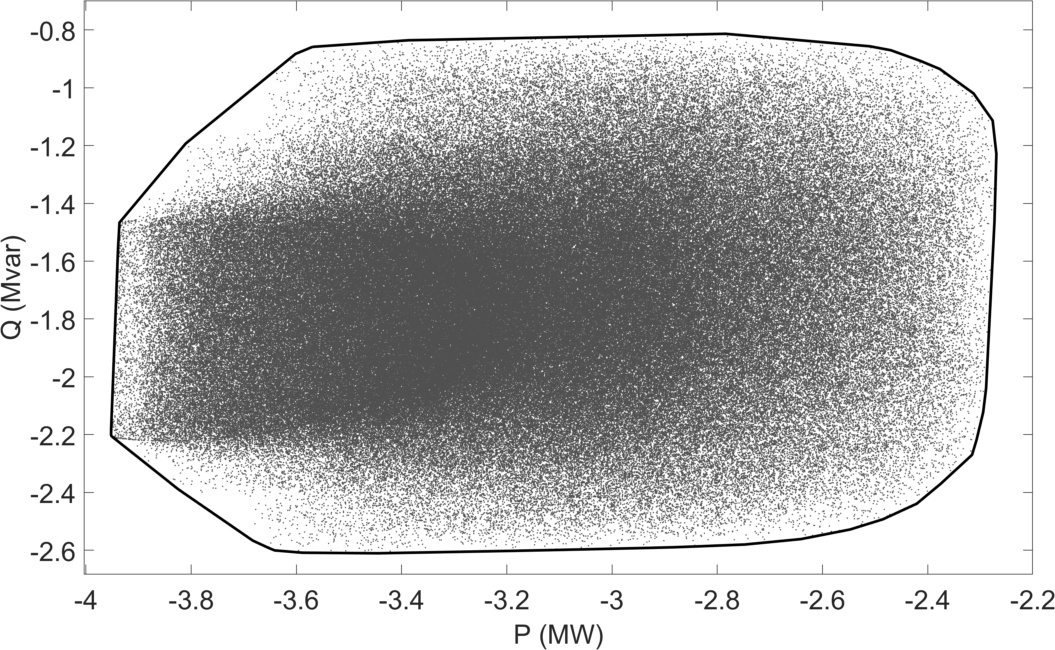}
	\caption{FOR of the VPP in Case I.}
	\vspace{-2em}
	\label{fig:FOR}
\end{figure}

The FOR of the VPP in Case I is shown in Fig.~\ref{fig:FOR},          
along with the convex hull of FOR approximating its boundary. The figure represents the PQ capability of the VPP and identifies the dispatchable active and reactive power levels for a specific instant of time. Due to the time-varying nature of demand and resources (e.g. flexible loads, PV and battery systems) the FOR of the VPP changes continuously and thus needs to be evaluated at regular intervals. For example, in an Australian context, with \SI{30}{\minute} commitment and \SI{5}{\minute} dispatch market, FOR needs to be evaluated with the time resolution of 5 min. FOR is particularly effective to evaluate the VPP capacity to bid in the energy market. After being dispatched at a specific dispatch point, the maximum feasible deviation represents the VPP potential to participate in grid support services.

\begin{figure}[!t]
	\centering
	\includegraphics[width=80mm] {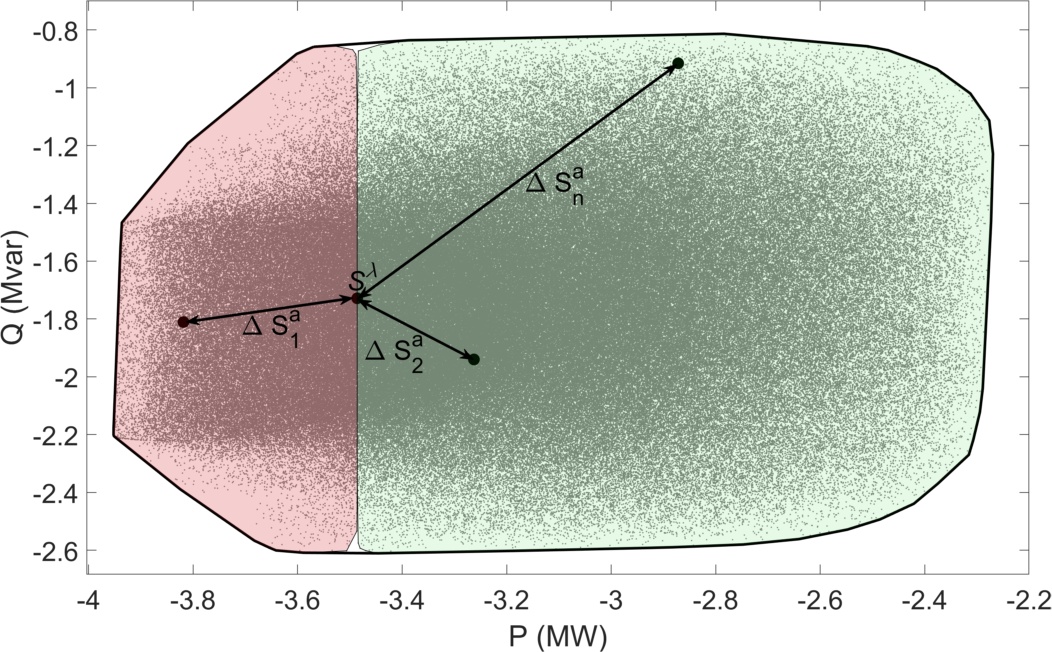}
	\caption{VPP feasibility regions (FOR) in Case I, along with the resulting potential regions for FCAS raise (green) and lower (red) services.}
	\label{fig:FOR_D}
\end{figure}

\textbf{Example:} Let’s assume that the VPP is arbitrarily dispatched by the energy market at point $S^{\lambda}$, and the complex power difference between any other point in the FOR and the dispatch point ($S^{\lambda}$) is called \textcolor{black}{ancillary power} ($\Delta S^\text{a}$). The \textcolor{black}{ancillary power} points with positive active power can provide FCAS raise\footnote{Raise FCAS are deployed to regulate, arrest, stabilise and recover drop in frequency.} services and points with negative active power can provide FCAS lower\footnote{Lower FCAS are deployed to regulate, arrest, stabilise and recover rise in frequency.} services, as illustrated in Fig.~\ref{fig:FOR_D}. However, in order to participate in various FCAS services (e.g. \SI{6}{\sec}, \SI{60}{\sec} or \SI{5}{\minute}~\cite{AEMO2015}) it is important to understand the time requirement associated with each \textcolor{black}{ancillary power point}.

\subsection{Flexibility}
\begin{figure}[!t]
	\centering
	\includegraphics[width=85mm] {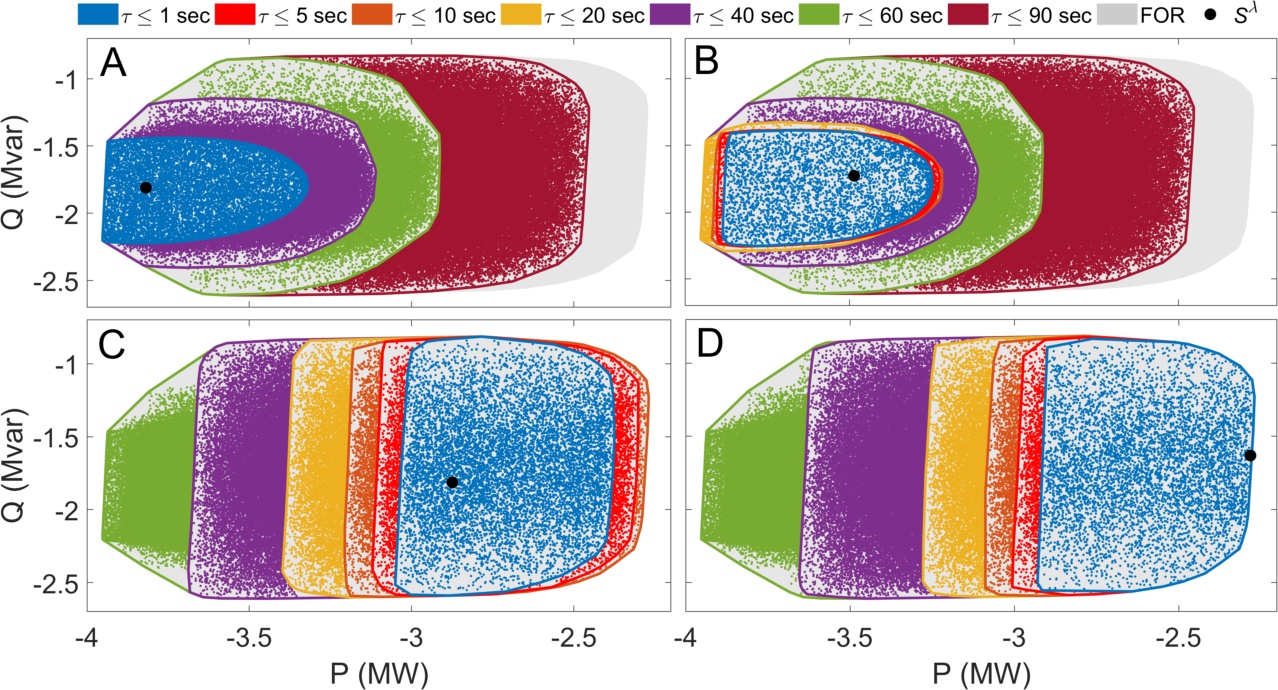}
	\caption{VPP flexibility regions (FXOR) in Case I, for four dispatch points.}
	\vspace{-1.5em}
	\label{fig:FLX}
\end{figure}

Flexibility is the quantitative measure of the time required to achieve \textcolor{black}{ancillary power point} from the dispatch power point. It depends upon factors such as status, activation and ramp time of the resources. It is particularly important in the presence of slow resources (e.g., diesel generators), as they could restrict the participation of the VPP in grid support services. The flexibility operating regions (FXOR) of the VPP in Case I are shown in Fig.~\ref{fig:FLX},
for four different dispatch points, characterising the strong interdependence between dispatch and flexibility. The dispatch point in Fig.~\ref{fig:FLX}-B and Fig.~\ref{fig:FOR_D} is the same, reflecting the additional information provided by the FXOR. While the results are mostly self-explanatory, Fig.~\ref{fig:FLX}-A is particularly interesting due to the large jump in the FXOR. The figure, in fact, shows that the \textcolor{black}{ancillary power} points in the blue area can be achieved within one second; however, no other point in the map is achievable under \SI{20}{\sec}, attributed to the fact that at this dispatch point all diesel generators are offline and the jump results due to the activation time requirement\footnote{Similar discontinuities in the operation of devices for the provision of ancillary services are also presented in~\cite{Mancarella2013C}}. In summary, a FXOR representation is a very useful tool for understanding VPP’s restriction to participate in FCAS markets because of the activation and ramp time constraints of the resources.
\subsubsection{Australian Context}
\begin{figure}[!t]
	\centering
	\includegraphics[width=80mm] {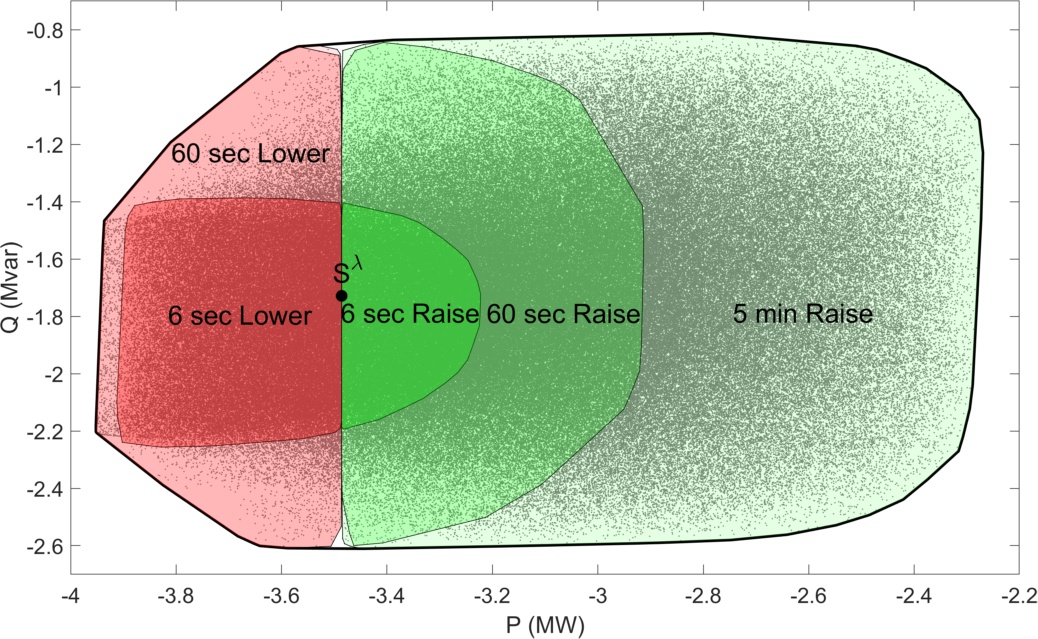}
	\vspace{-.5em}
	\caption{Case I VPP's potential for FCAS participation resulting from the dispatch point $S^\lambda$.}
	\label{fig:FLX_D}
\end{figure}

The Australian National Electricity Market (NEM) operates with six FCAS contingency services, namely,  Fast (\SI{6}{\sec}) Raise  and Lower, Slow (\SI{60}{\sec}) Raise and Lower, and Delayed (\SI{5}{\min}) Raise and Lower. Let us assume that the VPP is dispatched at $S^\lambda$ (same as depicted in Fig.~\ref{fig:FOR_D} and Fig.~\ref{fig:FLX}-B). Then, utilising the FXOR representation, the potential of VPP to participate in various FCAS markets can be identified, as shown in Fig.~\ref{fig:FLX_D}.        
\subsection{Impact of Flexible Demand and Network Topology} 
Additional flexible resources in a VPP will increase  its flexibility and results in larger FOR, along with a slight change in shape depending upon the PQ capabilities of the resources. If network constraints are not binding, then the resulting FOR can be accurately calculated using the VPP FOR and PQ charts of new resources. However, the impact of altering network topology is complicated and mainly depends upon the line limits and structure of the network.

\begin{figure}[!t]
	\centering
	\includegraphics[width=75mm] {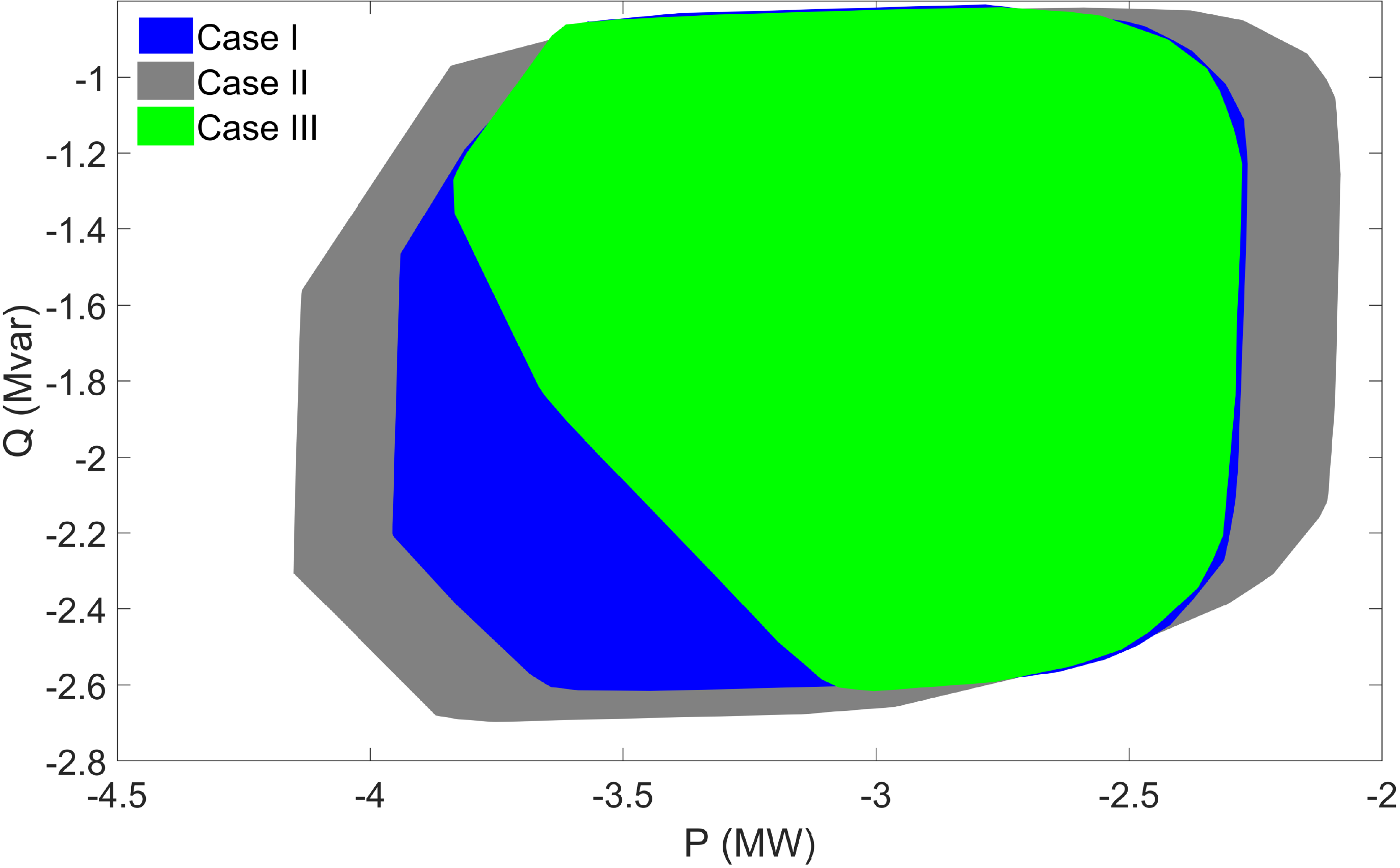}
	\caption{FORs for all three study cases.}
	\vspace{-1.5em}
	\label{fig:FOR_3}
\end{figure}

The FORs for all three study cases are shown in Fig.~\ref{fig:FOR_3}, 
representing that addition of flexible demand in Case II increases the FOR of the VPP, because of the direct relationship between FOR area and quantity of flexible resources and reconfiguration of the network by opening tie switches in Case III resulting into voltage limit violations for higher demand requirement of VPP thus eliminating corresponding points in FOR and FXOR.. In other words, network reconfiguration can reduce or enhance the FOR of VPP by affecting the operation of its resources due to underlying network constraints. 

\begin{figure}[!t]
	\centering
	\includegraphics[width=75mm] {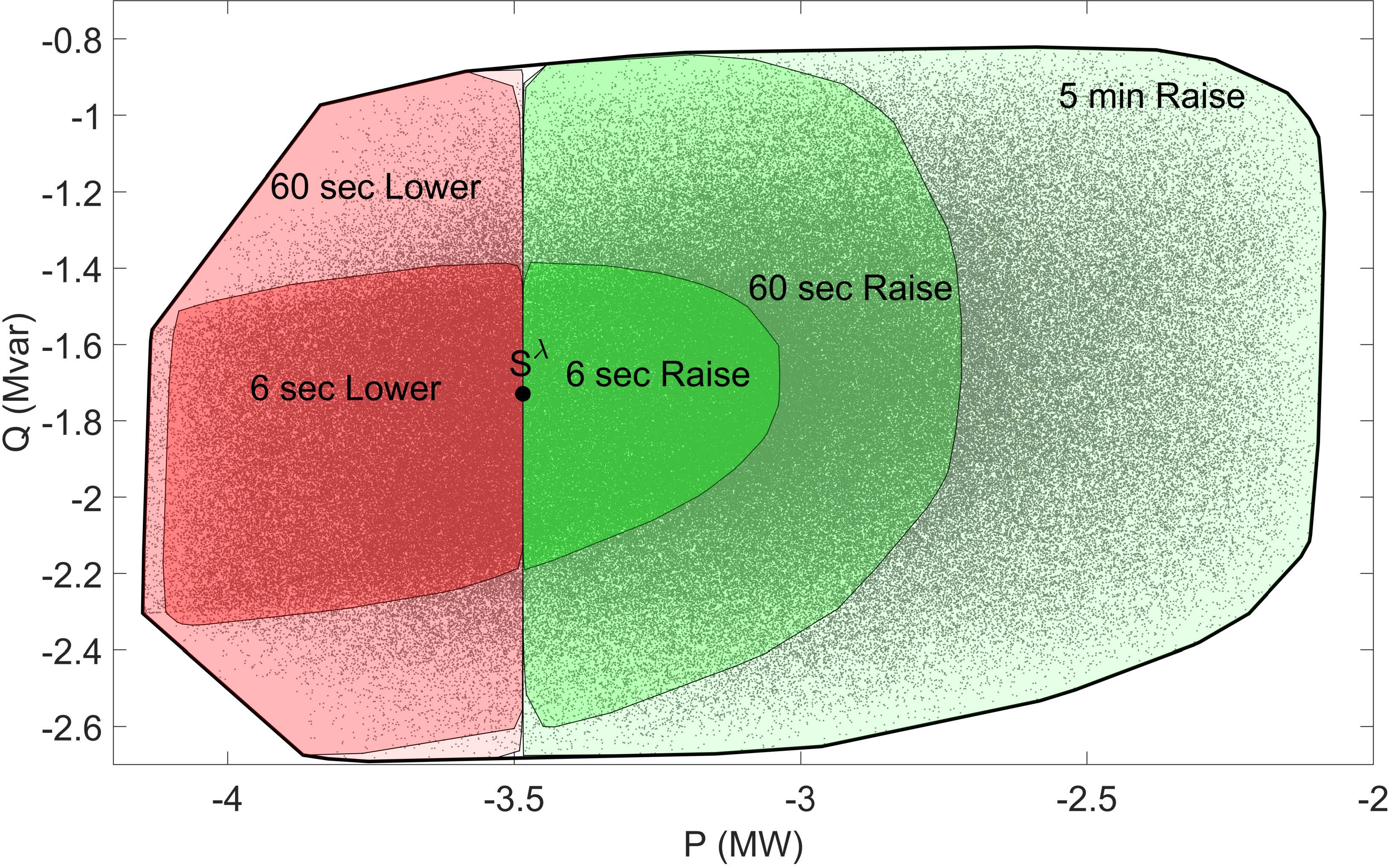}
	\vspace{-.5em}
	\caption{FCAS participation potential of VPP for Case II.}
	\label{fig:FLX_D2}
\end{figure}

\begin{figure}[!t]
	\centering
	\includegraphics[width=75mm] {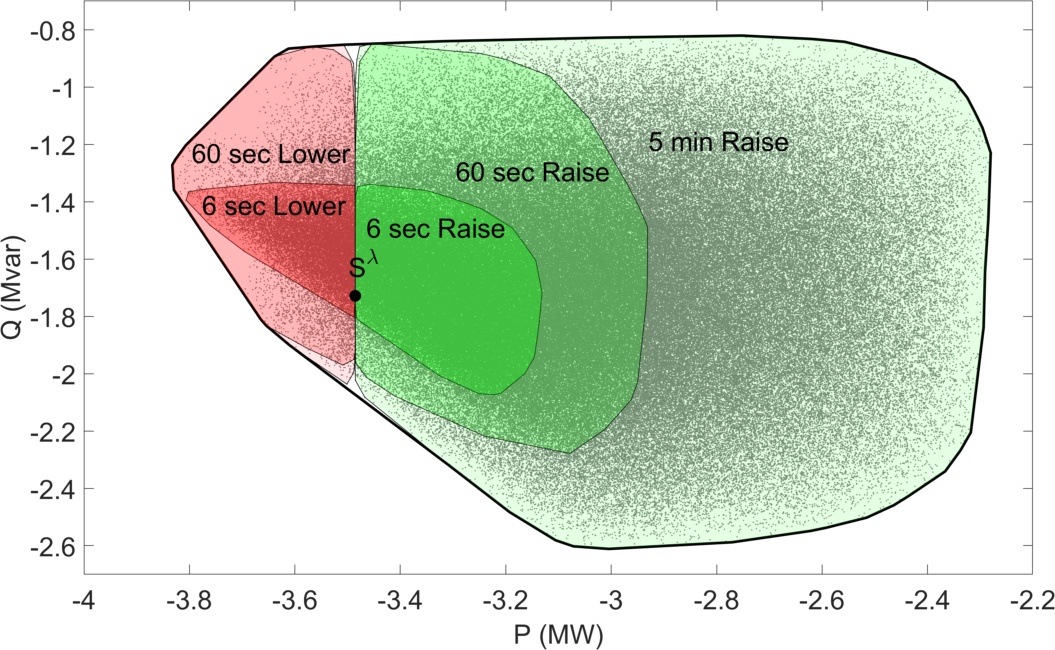}
	\vspace{-0.5em}
	\caption{FCAS participation potential of VPP for Case III.}
	\vspace{-1.5em}
	\label{fig:FLX_D3}
\end{figure}

In terms of flexibility, the addition of flexible demand which can be turned on or off rapidly increases the sub-second flexibility for FCAS participation, as represented by Fig.~\ref{fig:FLX_D2}.
On the other hand, Fig.~\ref{fig:FLX_D3} shows that VPP participation in all FCAS markets is affected by the network reconfiguration, which adds binding constraints mainly based on resource location and system demand.
Whereas, Fig.~\ref{fig:FLX_D3} represents the that the VPP participation factor in all contingency FCAS markets is effected due to the network restriction, which add binding constraints mainly based on resource location and system power demand. 

\section{Conclusion and Future Works} \label{Cnl}
This paper discusses how feasibility and flexibility for aggregation of DER in a VPP (or similarly for the purpose of modelling TSO/DSO interface operation) are not the same, with each concept bringing distinct useful information for the operation of VPP partaking in energy and grid support services. The feasibility space is characterized by the PQ capability chart of a VPP, whereas flexibility quantifies the FCAS (along with associated reactive power response capability) potential given a specific dispatch point. Two algorithms have been introduced to quantify the feasibility (FOR) and flexibility (FXOR) operating regions of a VPP. The case studies demonstrate the efficacy of the proposed methodology and algorithms and illustrate the impact of VPP’s dispatch on its FCAS participation. The case studies also reveal that the greater is the pool of resources (for example, adding DR), the larger will be the feasibility and flexibility potential of a VPP. However, the topology of the network is also an important parameter that introduces location-based constraints, which demands a further investigation to optimise the location of resources within a VPP.

\bibliographystyle{IEEEtran}
\bibliography{PowerTech2019V1}
\end{document}